\documentclass[11pt]{article}
\usepackage{jcappub,natbib,braket}
\input{colordvi.tex}
\title{
Signatures of anisotropic sources in the trispectrum of the cosmic
microwave background
}
\author[a,b]{Maresuke Shiraishi,}
\author[c,d]{Eiichiro Komatsu}
\author[e]{and Marco Peloso}
\affiliation[a]{Dipartimento di Fisica e Astronomia ``G. Galilei'', Universit\`a degli Studi di Padova, via Marzolo 8, I-35131, Padova, Italy}
\affiliation[b]{INFN, Sezione di Padova, via Marzolo 8, I-35131, Padova, Italy}
\affiliation[c]{Max-Planck-Institut f\"{u}r Astrophysik,
Karl-Schwarzschild Str. 1, 85741 Garching, Germany}
\affiliation[d]{Kavli Institute for the Physics and
Mathematics of the Universe, Todai Institutes for Advanced Study, the
University of Tokyo, Kashiwa, Japan 277-8583 (Kavli IPMU, WPI)}
\affiliation[e]{School of Physics and Astronomy, University of Minnesota, Minneapolis 55455, USA}

\def\ga{\mathrel{\raise.3ex\hbox{$>$\kern-.75em\lower1ex\hbox{$\sim$}}}}
\def\la{\mathrel{\raise.3ex\hbox{$<$\kern-.75em\lower1ex\hbox{$\sim$}}}}

\abstract{%
Soft limits of $N$-point correlation functions, in which one wavenumber is
much smaller than the others, play a special role in 
constraining the physics of inflation. Anisotropic sources such as a vector field during inflation generate distinct angular dependence in all these correlators. In this paper we focus on the four-point correlator (the trispectrum $T$). We adopt a parametrization motivated  by models in which the inflaton $\phi$ is coupled to a vector field through a $I^2 \left( \phi \right) F^2$ interaction, namely 
$T_{\zeta}({\bf k}_1, {\bf k}_2, {\bf k}_3, {\bf k}_4) 
\equiv \sum_n d_n 
[ P_n(\hat{\bf k}_1 \cdot \hat{\bf k}_3) 
+ P_n(\hat{\bf k}_1 \cdot \hat{\bf k}_{12}) 
+ P_n(\hat{\bf k}_3 \cdot \hat{\bf k}_{12}) 
] P_{\zeta}(k_1) P_{\zeta}(k_3) P_\zeta(k_{12})
  + (23~{\rm perm})$, 
where $P_n$ denotes the Legendre polynomials. This shape is enhanced when the wavenumbers of the diagonals of the quadrilateral are much smaller than the
sides, ${\bf k}_i$. The coefficient of the isotropic part, $d_0$, is 
equal to $\tau_{\rm NL}/6$ discussed in the literature. A  $I^2 \left( \phi \right) F^2$ interaction generates $d_2=2d_0$
which is, in turn, related to the quadrupole modulation parameter of the 
power spectrum, $g_*$, as $d_2\approx 14|g_*|N^2$ with $N\approx 60$. We show
that $d_0$ and $d_2$ can be equally well-constrained: the 
expected 68\%~CL error bars on these coefficients from a
cosmic-variance-limited experiment measuring temperature anisotropy of
the cosmic microwave background up to $\ell_{\rm max}=2000$ are $\delta
d_2 \approx 4 \delta d_0 = 105$. Therefore, we can reach $|g_*|=10^{-3}$
by measuring the angle-dependent trispectrum. The current upper limit on $\tau_{\rm NL}$ from the {\it Planck} temperature maps yields $|g_*|<0.02$ (95\%~CL).}

\begin{document}

\begin{flushright}  UMN-TH-3317/13 \end{flushright}

\maketitle
\flushbottom

\section{Introduction}

Cosmic inflation
\cite{Starobinsky:1980te,Sato:1980yn,Guth:1980zm,Linde:1981mu,Albrecht:1982wi}
is thought to have occurred in {\it nearly} de Sitter spacetime. Recent
convincing detection of a small deviation from the exact scale invariance of
primordial curvature perturbations \cite{Hinshaw:2012aka,Ade:2013lta}
shows that time-translation invariance is slightly broken during inflation. This
provides strong evidence for inflation, as the expansion rate during
inflation must be time-dependent in order for inflation to end
eventually, and the time dependence must be weak in order for inflation
to occur. This then leads to a natural question: ``Are other
symmetries also broken?'' 

Invariance under spatial rotation remains unbroken in the usual inflation models
based on scalar fields; however, it can be broken in the presence of
vector fields (see
ref.~\cite{Dimastrogiovanni:2010sm,Maleknejad:2012fw,Soda:2012zm} for
reviews). In such a case, the two-point correlation function in Fourier
space (power spectrum) of primordial curvature
perturbations defined by $\langle\zeta_{{\bf k}_1}\zeta_{{\bf
k}_2}\rangle=(2\pi)^3\delta^{(3)}({\bf k}_1+{\bf k}_2)P_\zeta({\bf
k}_1)$ generically exhibits a direction dependence as \cite{Ackerman:2007nb}
\begin{eqnarray}
P_\zeta({\bf k}) = P_0(k)\left[1 + g_*(k) (\hat{\bf k} \cdot \hat{\bf
			  E}_{\rm cl} )^2 \right],
\label{eq:gstar}
\end{eqnarray}
where $\hat{\bf E}_{\rm cl}$ is a preferred direction in space and
$P_0(k)$ is the isotropic power spectrum. The amplitude, $g_*(k)$, may
depend on wavenumbers. 

Temperature anisotropy of the cosmic microwave
background (CMB) offers a stringent test of rotational invariance of
correlation functions. Assuming that $g_*$ is independent of wavenumbers
(which is a reasonable assumption for inflation models we mostly focus
on in this paper, up to a logarithmic correction), ref.~\cite{Kim:2013gka} finds
$g_*=0.002\pm 0.016$ (68\%~CL) from the temperature data obtained
recently by the Planck satellite \cite{Ade:2013ktc}. The 95\%~CL limit is $-0.030<g_*<0.034$. This measurement
was achieved after removing statistical anisotropy caused by elliptical
beams of the {\it Planck} satellite \cite{Ade:2013dta} and emission from our
own Galaxy \cite{Ade:2013hta}. 

The three-point function (bispectrum) offers an additional test of 
rotational invariance of correlation functions. As breaking of
rotational invariance  
during inflation requires multiple fields (e.g., a scalar field driving
inflation and a vector field), it also breaks the so-called
single-field consistency relation of the bispectrum
\cite{Maldacena:2002vr,Creminelli:2004yq}; namely, there can be a
non-negligible correlation in a ``soft limit'' of the three-point
correlation function, in which one wavenumber, say $k_3$, is much
smaller than the other two, i.e., $k_3\ll k_1\approx k_2$. Breaking of
rotational invariance then introduces a dependence of the soft-limit
bispectrum on angles between the wavenumbers. Defining the bispectrum as
$\langle\zeta_{{\bf k}_1}\zeta_{{\bf k}_2}\zeta_{{\bf
k}_3}\rangle=(2\pi)^3\delta^{(3)}({\bf k}_1+{\bf k}_2+{\bf
k}_3)B_\zeta(k_1,k_2,k_3)$, we write \cite{Shiraishi:2013vja}
\begin{eqnarray}
B_\zeta (k_1,k_2,k_3) 
= \sum_n c_n P_n(\hat{\bf k}_1 \cdot \hat{\bf
 k}_2)P_\zeta(k_1)P_\zeta(k_2)+(\mbox{2 perm}) \label{eq:zeta_bis_def},
\end{eqnarray}
where $P_n(x)$ denotes the Legendre polynomials. Note that this form is
valid for an isotropic measurement of the bispectrum, namely for the case in which we fix a triangular shape, and we then average over all possible orientations of this shape in Fourier space (this is equivalent to taking an average over all possible directions for the preferred direction $\hat{\bf E}_{\rm cl}$). 
The {\it Planck} temperature data give constraints on the first three
coefficients as $c_0=3.2\pm 7.0$, $c_1 = 11.0 \pm 113$, and $c_2 = 3.8
\pm 27.8$ ($68\%$~CL) \cite{Ade:2013ydc}. Given a model of inflation,
these coefficients can be related to the parameter in the power
spectrum, $g_*$. For example, the relation is
$c_0=2c_2=320|g_*|(N/60)$ (with $N\approx 60$ being the number of
$e$-folds counted from the end of inflation) for 
inflation models with a scalar field driving
inflation, $\phi$, coupled to a vector field in the form of $I^2(\phi)F^2$
where $F$ is a vector-field strength tensor
\cite{Barnaby:2012tk,Bartolo:2012sd,Shiraishi:2013vja,Abolhasani:2013zya, Fujita:2013qxa}.
\footnote{A bispectrum with a nontrivial angular dependence in the squeezed limit is also obtained in the model of solid inflation \cite{Endlich:2012pz}, which is a model characterized by three scalar fields with a nontrivial spatial profile. In this model, $c_2 \gg c_0$.}
We then obtain $|g_*|<0.05$ and $0.36$ (95\%~CL) from $c_0$ and
$c_2$, respectively.

The goal of this paper is to investigate the four-point function
(trispectrum) defined by $\Braket{\prod_{i=1}^4 \zeta_{{\bf k}_i}}_c =
(2\pi)^3 T_{\zeta}({\bf k}_1, {\bf k}_2, {\bf k}_3, {\bf k}_4)
\delta^{(3)} \left( \sum_{i=1}^4 {\bf k}_i \right)$, 
where $\langle\dots\rangle_c$ denotes the connected part of the
trispectrum. 
The trispectrum is fully parametrized by six independent numbers,
i.e., three wavenumbers and three angles between wavevectors, e.g.,
$k_1$, $k_3$, $k_{12}$, $\hat{\bf k}_1 \cdot \hat{\bf k}_3$, $\hat{\bf
k}_1 \cdot \hat{\bf k}_{12}$, and $\hat{\bf k}_3 \cdot \hat{\bf
k}_{12}$, where ${\bf k}_{12} \equiv {\bf k}_{1} + {\bf k}_2$.
\footnote{With this parametrization, we divide the quadrilateral in the two triangles having sides $k_1$, $k_2$, $k_{12}$, and $k_3$, $k_4$, $k_{12}$, respectively. We then specify each triangle and their relative orientation.} 
We then find a simple linear parametrization as  
\begin{eqnarray}
T_\zeta({\bf k}_1, {\bf k}_2, {\bf k}_3, {\bf k}_4) 
&=& \sum_n 
\left[ A_n P_n(\hat{\bf k}_1 \cdot \hat{\bf k}_3) + B_n P_n(\hat{\bf k}_1 \cdot \hat{\bf k}_{12}) + C_n P_n(\hat{\bf k}_3 \cdot \hat{\bf k}_{12}) \right] \nonumber \\ 
&&\times P_{\zeta}(k_1) P_{\zeta}(k_3) P_\zeta(k_{12})  + (23~{\rm perm})~.
\end{eqnarray}
Symmetry under permutations of ${\bf k}_i$ imposes $B_n = C_n$, while $A_n$ remains independent in
general. By construction, this trispectrum has the largest values in
soft limits in which diagonals of a quadrilateral, $k_{12}$, etc., are
much smaller than the sides, $k_i$.  

Instead of studying the most general form, we shall study a
simpler form motivated by inflation with $I^2(\phi)F^2$ coupling, which
yields $A_n = B_n = C_n$ \cite{Shiraishi:2013vja, Abolhasani:2013zya,
Fujita:2013qxa}. Our parametrization is
\begin{eqnarray}
T_{\zeta}({\bf k}_1, {\bf k}_2, {\bf k}_3, {\bf k}_4) 
&=& \sum_n d_n
\left[ P_n(\hat{\bf k}_1 \cdot \hat{\bf k}_3) + P_n(\hat{\bf k}_1 \cdot \hat{\bf k}_{12}) + P_n(\hat{\bf k}_3 \cdot \hat{\bf k}_{12}) \right] \nonumber \\ 
&&\times P_{\zeta}(k_1) P_{\zeta}(k_3) P_\zeta(k_{12})  + (23~{\rm perm})~. \label{eq:zeta_tris_def}
\end{eqnarray} 
The readers who are familiar with
the primordial trispectrum would find that the first coefficient, $d_0$,
is equal to $\tau_{\rm NL}/6$ in the literature
\cite{Boubekeur:2005fj}. Again, this form is
valid for the trispectrum averaged over all possible directions of
quadrilaterals (or $\hat{\bf E}_{\rm cl}$). As we shall show later in
section~\ref{sec:g}, 
$I^2(\phi)F^2$ inflation gives $d_0=d_2/2\approx
7|g_*|N^2$.
\footnote{The odd terms in the expansion (\ref{eq:zeta_tris_def}) 
 may arise if the source of the anisotropic modulation breaks parity.} 
While $d_2$ is yet to be constrained by the data, the current
upper limit on $d_0=\tau_{\rm NL}/6<470$ (95\%~CL) from the {\it Planck} data
\cite{Ade:2013ydc} yields 
$|g_*|<0.02$ (95\%~CL), which is already better 
than the limit from the power spectrum or the bispectrum. The
limits on $d_0$ and $d_2$ from the next {\it Planck} data release should
improve the limit further. 

This paper is organized as follows. In section~\ref{sec:cmbtri}, we
calculate the trispectrum of CMB temperature anisotropy from
eq.~\eqref{eq:zeta_tris_def} both with the flat-sky approximation and
the full-sky formalism. In section~\ref{sec:Fisher}, we calculate the
expected 68\%~CL error bars on $d_0$ and $d_2$ from a
cosmic-variance-limited CMB experiment. In section~\ref{sec:g}, we
translate the error bars on $d_0$ and $d_2$ to that on $g_*$. We
conclude in section~\ref{sec:conclusion}. 

\section{Trispectrum of CMB temperature anisotropy}
\label{sec:cmbtri}

Let us rewrite the trispectrum given in eq.~\eqref{eq:zeta_tris_def} as 
\begin{eqnarray}
\Braket{\prod_{i=1}^4 \zeta_{{\bf k}_i}}_c &=& (2\pi)^3 \int d^3 {\bf K} \delta^{(3)}\left({\bf k}_1 + {\bf k}_2 + {\bf K} \right) \delta^{(3)} \left({\bf k}_3 + {\bf k}_4 - {\bf K} \right) 
\sum_n d_n 
{t}^{{\bf k}_1 {\bf k}_2}_{{\bf k}_3 {\bf k}_4} ({\bf K},n) \nonumber \\ 
&& + (23~{\rm perm})~, 
\end{eqnarray}
with
\begin{eqnarray}
{t}_{{\bf k}_3 {\bf k}_4}^{{\bf k}_1{\bf k}_2} ({\bf K},n) 
 &\equiv& 
 \left[ P_n(\hat{\bf k}_1 \cdot \hat{\bf k}_3) 
+ \frac{1 + (-1)^n}{2} P_n(\hat{\bf k}_1 \cdot \hat{\bf K}) 
+ \frac{1 + (-1)^n}{2} P_n(\hat{\bf k}_3 \cdot \hat{\bf K}) \right] \nonumber \\ 
&&\times P_\zeta(k_1) P_\zeta(k_3) P_\zeta(K)~. \label{eq:zeta_tris_CMB}
\end{eqnarray}
Here, a {\it reduced} curvature trispectrum, ${t}_{{\bf k}_3 {\bf
k}_4}^{{\bf k}_1{\bf k}_2} ({\bf K},n)$, satisfies ${t}_{{\bf k}_3 {\bf
k}_4}^{{\bf k}_1{\bf k}_2} ({\bf K},n) = {t}^{{\bf k}_3 {\bf k}_4}_{{\bf
k}_1{\bf k}_2} ({\bf K},n)$.  
We then write eq.~\eqref{eq:zeta_tris_CMB} using spherical harmonics as
\begin{eqnarray}
{t}_{{\bf k}_3 {\bf k}_4}^{{\bf k}_1{\bf k}_2} ({\bf K}, n) 
&=& P_\zeta(k_1) P_\zeta(k_3) P_\zeta(K) 
\frac{4\pi}{2n+1} \sum_{\mu} 
\nonumber \\ 
&&\times 
\left[ Y_{n \mu}^* (\hat{\bf k}_1) Y_{n \mu} (\hat{\bf k}_3) 
+ \frac{1 + (-1)^n}{2} 
\left( Y_{n \mu}^* (\hat{\bf k}_1) + Y_{n \mu}^* (\hat{\bf k}_3) \right) 
Y_{n \mu} (\hat{\bf K})
\right]  ~. \label{eq:zeta_tris_Ylm}
\end{eqnarray}

\subsection{Flat-sky formula}

To gain analytical insights into the structure of the CMB trispectrum,
we first derive the CMB trispectrum in the flat-sky approximation. The
coefficients of the two-dimensional Fourier transform of temperature
anisotropy in a small flat section of the sky are related to the
curvature perturbation as \cite{Shiraishi:2010sm}
\begin{eqnarray}
a({\boldsymbol \ell}) &=& \int_0^{\tau_0} d\tau \int_{-\infty}^{\infty}
 \frac{dk_z}{2 \pi} \zeta \left({\bf k}^{\parallel} 
= -\frac{\boldsymbol \ell}{D}, k_z\right) 
 S_I\left(k = \sqrt{k_z^2 + (\ell/D)^2},\tau\right) \frac{1}{D^2}e^{- i k_z D} ~,
\end{eqnarray}
where $D \equiv \tau_0 - \tau$ denotes the conformal distance between a
given conformal time, $\tau$, and the present time, $\tau_0$; ${\bf k} =
({\bf k}^\parallel, k_z)$ with ${\bf k}^\parallel=(k_x,k_y)$; and $S_I$
is the so-called source function. The flat-sky approximation is accurate
for $\ell \gg 1$. 

The trispectrum of $a({\boldsymbol \ell})$ in the limits of $\ell_i \gg k_{i} D$ and $L \gg k_z r$ is given by
\begin{eqnarray}
\Braket{\prod_{i=1}^4 a({\boldsymbol \ell}_i) }_c
&=& (2\pi)^2
\int d^2 {\bf L} 
\delta^{(2)}\left(\boldsymbol{\ell}_1 + \boldsymbol{\ell}_2 + {\bf L} \right) 
\delta^{(2)}\left(\boldsymbol{\ell}_3 + \boldsymbol{\ell}_4 - {\bf L} \right) 
\sum_{n} d_n t^{\boldsymbol{\ell}_1\boldsymbol{\ell}_2}_{\boldsymbol{\ell}_3\boldsymbol{\ell}_4}({\bf L},n) \nonumber \\ 
&&+ (23~{\rm perm}) ~,
\end{eqnarray}
where
$t^{\boldsymbol{\ell}_1\boldsymbol{\ell}_2}_{\boldsymbol{\ell}_3\boldsymbol{\ell}_4}({\bf
L}, n)$ is the so-called CMB reduced trispectrum:
\begin{eqnarray}
t^{\boldsymbol{\ell}_1\boldsymbol{\ell}_2}_{\boldsymbol{\ell}_3\boldsymbol{\ell}_4}({\bf L},n) &\equiv& 
\int_{-\infty}^\infty r^2 dr 
\left[ \prod_{i=1}^4 \int_0^{\tau_0} d\tau_i \int_{\ell_i / D_i}^{\infty}
\frac{dk_{i}}{2 \pi} {\cal G}(\ell_i, k_i, \tau_i, r) \right] 
P_\zeta(k_1) P_\zeta(k_3) P_\zeta\left(\frac{L}{|r|} \right) 
 \nonumber \\ 
&& \times 
 \left[ P_n( \hat{\boldsymbol{\ell}}_1 \cdot \hat{\boldsymbol{\ell}}_3 ) 
+ \frac{1 + (-1)^n}{2} P_n( \hat{\boldsymbol{\ell}}_1 \cdot \hat{\bf L}) 
+ \frac{1 + (-1)^n}{2} P_n ( \hat{\boldsymbol{\ell}}_3 \cdot \hat{\bf L} ) 
\right], \label{eq:CMB_tris_flat}
\end{eqnarray} 
with
\begin{eqnarray}
{\cal G}(\ell, k, \tau, r)
= \left[1 - \left(\frac{\ell}{kD}\right)^2 \right]^{-1/2}
 S_I\left(k,\tau \right) \frac{2}{D^2} 
\cos\left[\sqrt{1 - \left(\frac{\ell}{kD}\right)^2} k( r - D)\right]~. 
\end{eqnarray}
The flat-sky reduced CMB trispectrum directly reflects the angular
dependence of the Legendre 
polynomials in the reduced curvature trispectrum, ${t}_{{\bf k}_3 {\bf
k}_4}^{{\bf k}_1{\bf k}_2} ({\bf K},n)$, given in eq.~(\ref{eq:zeta_tris_CMB}).  

\begin{figure}
  \begin{center}
    \includegraphics[width = 0.75\textwidth]{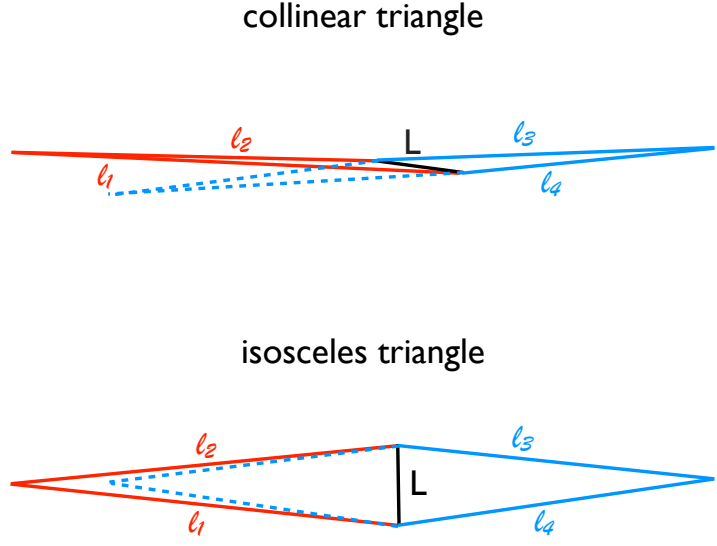}
  \end{center}
  \caption{Two shapes of the trispectrum in $\ell$ space in soft limits
 in which the diagonal, $L$, is much smaller than the sides,
 $\ell_i$. The red lines show $\ell_1$ and $\ell_2$, while the blue
 solid and dashed lines show two configurations of $\ell_3$ and
 $\ell_4$. Specifically, the blue solid and dashed lines show
 configurations in which 
 $\hat{\boldsymbol{\ell}}_1 \cdot \hat{\boldsymbol{\ell}}_3 \approx - 1$
 and $1$, respectively. The black lines show the diagonal, $L$.
(Top) Collinear configurations: $\hat{\boldsymbol{\ell}}_1
 \cdot \hat{\bf L} \approx -1$, $\hat{\boldsymbol{\ell}}_3 \cdot
 \hat{\bf L} \approx \mp 1 $, and $\hat{\boldsymbol{\ell}}_1 \cdot
 \hat{\boldsymbol{\ell}}_3 \approx \pm 1$. (Bottom) Isosceles
 configurations: $\hat{\boldsymbol{\ell}}_1 \cdot \hat{\bf L} \approx \hat{\boldsymbol{\ell}}_3 \cdot \hat{\bf L} \approx 0$, and $\hat{\boldsymbol{\ell}}_1 \cdot \hat{\boldsymbol{\ell}}_3 \approx \pm 1$.}
  \label{fig:triangle}
\end{figure}

The isotropic term, $n=0$, has the largest values when the diagonal,
$L$, is much smaller than the sides, $\ell_i$, i.e.,
$\ell_1 \approx \ell_2 \gg L$ or $\ell_3 \approx \ell_4 \gg L$
\cite{Kogo:2006kh}. The amplitude of the trispectrum in this limit is modulated when $n\neq 0$. For example, in the ``collinear
configurations,'' $\hat{\boldsymbol{\ell}}_1 \cdot \hat{\bf L}
\approx -1$, $\hat{\boldsymbol{\ell}}_3 \cdot \hat{\bf L} \approx \mp 1
$, and $\hat{\boldsymbol{\ell}}_1 \cdot \hat{\boldsymbol{\ell}}_3
\approx \pm 1$ (see the top panel of figure~\ref{fig:triangle}), 
we find 
\footnote{We have the same relationship between magnitudes for the other
collinear configurations: $\hat{\boldsymbol{\ell}}_1 \cdot \hat{\bf L}
\approx 1$, $\hat{\boldsymbol{\ell}}_3 \cdot \hat{\bf L} \approx \pm 1
$, and $\hat{\boldsymbol{\ell}}_1 \cdot \hat{\boldsymbol{\ell}}_3
\approx \pm 1$.}
\begin{eqnarray}
t^{\boldsymbol{\ell}_1\boldsymbol{\ell}_2}_{\boldsymbol{\ell}_3\boldsymbol{\ell}_4}({\bf L},0) : 
t^{\boldsymbol{\ell}_1\boldsymbol{\ell}_2}_{\boldsymbol{\ell}_3\boldsymbol{\ell}_4}({\bf L},1) : 
t^{\boldsymbol{\ell}_1\boldsymbol{\ell}_2}_{\boldsymbol{\ell}_3\boldsymbol{\ell}_4}({\bf 
L},2) 
\approx 1 : \pm \frac{1}{3} : 1 ~. \label{eq:CMB_tris_flat_collinear}
\end{eqnarray}
In the ``isosceles configurations,'' $\hat{\boldsymbol{\ell}}_1 \cdot
\hat{\bf L} \approx \hat{\boldsymbol{\ell}}_3 \cdot \hat{\bf L} \approx
0$, and $\hat{\boldsymbol{\ell}}_1 \cdot \hat{\boldsymbol{\ell}}_3
\approx \pm 1$ (see the bottom panel of figure~\ref{fig:triangle}), we
find that the $n=2$ trispectrum vanishes:
\begin{eqnarray}
t^{\boldsymbol{\ell}_1\boldsymbol{\ell}_2}_{\boldsymbol{\ell}_3\boldsymbol{\ell}_4}({\bf L},0) : 
t^{\boldsymbol{\ell}_1\boldsymbol{\ell}_2}_{\boldsymbol{\ell}_3\boldsymbol{\ell}_4}({\bf L},1) : 
t^{\boldsymbol{\ell}_1\boldsymbol{\ell}_2}_{\boldsymbol{\ell}_3\boldsymbol{\ell}_4}({\bf 
L},2) 
\approx 1 : \pm \frac{1}{3} : 0 ~. \label{eq:CMB_tris_flat_isosceles}
\end{eqnarray}
Note that the sign of the $n=1$ trispectrum can change, as the Legendre
polynomial with $n=1$ is an odd function. These signatures will affect
the expected error bars on $d_n$ as discussed in section~\ref{sec:Fisher}. 

\subsection{Full-sky formula}

We shall move onto the full-sky formalism. The spherical harmonics
coefficients of temperature anisotropy are related to the curvature
perturbation as 
\begin{eqnarray}
a_{\ell m} = 
4\pi (-i)^{\ell} \int \frac{k^2 dk}{(2\pi)^3} {\cal T}_{\ell}(k) \zeta_{\ell m}(k),
\end{eqnarray}
where $\zeta_{\ell m}(k)$ is the curvature perturbation in spherical
harmonics space: $\zeta_{\ell m}(k) \equiv \int d^2 \hat{\bf k}
\zeta({\bf k}) Y_{\ell m}^* (\hat{\bf k})$, and ${\cal T}_{\ell}(k)$ is
the radiation transfer function, which is related to the source function
as ${\cal T}_\ell(k) = \int_0^{\tau_0} d\tau~S_I(k,\tau)
j_\ell(kD)$. Using this and the computational technique developed in
ref.~\cite{Shiraishi:2010kd}, the CMB trispectrum is given by 
\begin{eqnarray}
\Braket{\prod_{i=1}^4 a_{\ell_i m_i}}_c 
&=& \left[\prod_{i=1}^4 4\pi (-i)^{\ell_i} \int \frac{k_i^2 dk_i}{(2\pi)^3} {\cal T}_{\ell_i}(k_i) \right] \Braket{\prod_{i=1}^4 \zeta_{\ell_i m_i}(k_i)}_c ~, \label{eq:CMB_tris_def}
\end{eqnarray}
where 
\begin{eqnarray}
\Braket{\prod_{i=1}^4 \zeta_{\ell_i m_i}(k_i)}_c
&=&  \sum_{L M} (-1)^{M} \left(
  \begin{array}{ccc}
  \ell_1 & \ell_2 & L \\
  m_1 & m_2 & -M 
  \end{array}
 \right)
\left(
  \begin{array}{ccc}
  \ell_3 & \ell_4 & L \\
  m_3 & m_4 & M 
  \end{array}
 \right) \nonumber \\ 
&&\times 
(2\pi)^3 \sum_n d_n {t}_{k_3 k_4 \ell_3 \ell_4}^{k_1 k_2 \ell_1 \ell_2}(L,n) + (23 ~{\rm perm} ),
\end{eqnarray}
and
\begin{eqnarray}
{t}_{k_3 k_4 \ell_3 \ell_4}^{k_1 k_2 \ell_1 \ell_2}(L, n) 
&=& 
 P_\zeta(k_1) P_\zeta(k_3) 
\sum_{L_1 L_3 L'}  8^2 (-1)^{\frac{L_1+ \ell_2 +L_3 + \ell_4}{2} 
+ \ell_1 + \ell_2 + \ell_3 + \ell_4} 
 \nonumber \\ 
&&\times  
\int_0^\infty r^2 dr \int_0^\infty r'^2 dr' 
j_{L_1}(k_1 r) 
j_{\ell_2}(k_2 r)  j_{L_3}(k_3 r')  j_{\ell_4}(k_4 r') 
\nonumber \\ 
&&\times 
\frac{\pi}{2} \left[ 
F_{L' L'}(r, r') {\cal I}_{\ell_3 \ell_4}^{\ell_1 \ell_2}(L_1, L_3, L'; n, L)
+ F_{L' L}(r, r')  
{\cal J}_{\ell_3 \ell_4}^{\ell_1 \ell_2}(L_1, L_3, L'; n, L)
\right. \nonumber \\  
&& \left. \qquad  
+ F_{L L'}(r, r')  
{\cal J}^{\ell_3 \ell_4}_{\ell_1 \ell_2}(L_3, L_1, L'; n, L) 
\right] ~. \label{eq:zeta_red_tris}
 \end{eqnarray}
The $F$ function, defined by 
\begin{eqnarray}
F_{LL'}(r,r') &\equiv& \frac{2}{\pi}\int K^2 dK P_\zeta(K) j_L(Kr) j_{L'}(Kr')~, 
\end{eqnarray}
projects the $K$ dependence onto $L$. The ${\cal I}$ and ${\cal J}$
symbols are defined by
\begin{eqnarray}
{\cal I}_{\ell_3 \ell_4}^{\ell_1 \ell_2}(L_1, L_3, L'; n, L) 
&\equiv& \frac{4\pi}{2n+1} 
(-1)^{\ell_2 + \ell_4 + L' + n + L} 
I_{L_1 \ell_2 L'} 
 I_{L_3 \ell_4 L'}  
I_{\ell_1 L_1 n}I_{\ell_3 L_3 n} \nonumber \\ 
&&\times 
 (2L + 1)
\left\{
  \begin{array}{ccc}
  \ell_1 & \ell_2 & L \\
  L' & n & L_1 
  \end{array}
 \right\}
\left\{
  \begin{array}{ccc}
 \ell_3 & \ell_4 & L \\
  L' & n & L_3 
  \end{array}
 \right\} ~, \\ 
{\cal J}_{\ell_3 \ell_4}^{\ell_1 \ell_2}(L_1, L_3, L'; n, L)
&\equiv& \frac{4\pi}{2n+1}  \frac{1 + (-1)^n}{2} 
(-1)^{\ell_2 + L + \frac{L'+L}{2}} I_{L' L n} 
I_{L_1 \ell_2 L'} 
 I_{\ell_3 \ell_4 L} 
I_{\ell_1 L_1 n}  \nonumber \\ 
&&\times 
 \delta_{L_3, \ell_3}
\left\{
  \begin{array}{ccc}
 \ell_1 & \ell_2 & L \\
 L' & n & L_1 
  \end{array}
 \right\} ~,
\end{eqnarray} 
with $I_{l_1 l_2 l_3}
\equiv \sqrt{\frac{(2 l_1 + 1)(2 l_2 + 1)(2 l_3 + 1)}{4 \pi}}
\left(
  \begin{array}{ccc}
  l_1 & l_2 & l_3 \\
  0 & 0 & 0
  \end{array}
 \right)$, and they reflect characteristic $\ell$ dependence imposed by
 $P_n(\hat{\bf k}_1 \cdot \hat{\bf k}_3)$ and $P_n(\hat{\bf k}_{1,
 3} \cdot \hat{\bf K})$, respectively. The selection rules in these
 symbols restrict summation ranges of $L_1$, $L_3$ and $L'$ to the
 values close to $\ell_1$, $\ell_3$ and $L$, respectively. They also
 guarantee parity invariance of the trispectrum; namely, $\ell_1 +
 \ell_2 + \ell_3 + \ell_4 = {\rm even}$ although $\ell_1 + \ell_2$ or
 $\ell_3 + \ell_4$ can take on both even and odd numbers in the ${\cal
 I}$ function. Substituting eq.~(\ref{eq:zeta_red_tris}) into
 eq.~(\ref{eq:CMB_tris_def}) leads to the final expression of the CMB
 trispectrum: 
\begin{eqnarray}
\Braket{\prod_{i=1}^4 a_{\ell_i m_i}}_c
&=& \sum_{LM} (-1)^{M} 
 \left(
  \begin{array}{ccc}
  \ell_1 & \ell_2 & L \\
  m_1 & m_2 & -M 
  \end{array}
 \right)
\left(
  \begin{array}{ccc}
  \ell_3 & \ell_4 & L \\
  m_3 & m_4 & M 
  \end{array}
 \right) \nonumber \\
&&\times  \sum_n d_n t_{\ell_3 \ell_4}^{\ell_1 \ell_2}(L,n) + (23 ~{\rm perm})~,
\end{eqnarray} 
where the reduced form is given by
\begin{eqnarray}
 t_{\ell_3 \ell_4}^{\ell_1 \ell_2}(L,n) 
&\equiv&
\sum_{L_1 L_3 L'}  (-1)^{\frac{L_1 + L_3 + \ell_1 + \ell_3 }{2} + \ell_1 + \ell_3} 
 \nonumber \\ 
&&\times  
\int_0^\infty r^2 dr \int_0^\infty r'^2 dr' 
\beta_{\ell_1 L_1}(r) 
\alpha_{\ell_2}(r)  \beta_{\ell_3 L_3}(r') \alpha_{\ell_4}(r') 
\nonumber \\ 
&&\times 
\left[ 
F_{L' L'}(r, r') {\cal I}_{\ell_3 \ell_4}^{\ell_1 \ell_2}(L_1, L_3, L'; n, L)
+ F_{L' L}(r, r')  
{\cal J}_{\ell_3 \ell_4}^{\ell_1 \ell_2}(L_1, L_3, L'; n, L)
\right. \nonumber \\  
&& \left. \quad  
+ F_{L L'}(r, r')  
{\cal J}^{\ell_3 \ell_4}_{\ell_1 \ell_2}(L_3, L_1, L'; n, L) 
\right] ~, \label{eq:CMB_red_tris}  
\end{eqnarray}
with
\begin{eqnarray}
\alpha_{\ell}(r) &=& \frac{2}{\pi} \int k^2 dk {\cal T}_\ell(k) j_\ell (kr) ~, \\ 
\beta_{\ell L}(r) &=& \frac{2}{\pi} \int k^2 dk  P_\zeta(k) {\cal T}_\ell(k) j_L (kr) ~. 
\end{eqnarray}
When $n=0$, we have 
\begin{eqnarray}
{\cal I}_{\ell_3 \ell_4}^{\ell_1 \ell_2}(L_1, L_3, L'; 0, L) 
&=& {\cal J}_{\ell_3 \ell_4}^{\ell_1 \ell_2}(L_1, L_3, L'; 0, L) 
= {\cal J}^{\ell_3 \ell_4}_{\ell_1 \ell_2}(L_3, L_1, L'; 0, L) \nonumber \\  
&=&
I_{\ell_1 \ell_2 L} I_{\ell_3 \ell_4 L} \delta_{\ell_1, L_1} \delta_{\ell_3, L_3} \delta_{L' L}  ~,
\end{eqnarray}
which agrees with the previous results \cite{Okamoto:2002ik,Kogo:2006kh}.  

The $r$ and $r'$ integrals in eq.~(\ref{eq:CMB_red_tris}) are dominated
by contributions from $r \simeq r' \simeq r_* \equiv \tau_0 - \tau_*$
with $\tau_*$ being the recombination epoch, as $\alpha_\ell(r)$ and
$\beta_{\ell L}(r)$ peak at $r \simeq r_*$. If $F_{LL'}(r, r')$ varies
slowly for $r \simeq r' \simeq r_*$, i.e., in the small-$L$ limit, the
$r$ and $r'$ integrals may become separable:
\begin{eqnarray}
 t_{\ell_3 \ell_4}^{\ell_1 \ell_2}(L,n) 
&\approx& 
\sum_{L_1 L_3 L'}  (-1)^{\frac{L_1 + L_3 + \ell_1 + \ell_3 }{2} + \ell_1 + \ell_3} 
R(\ell_1, L_1, \ell_2) R(\ell_3, L_3, \ell_4)
\nonumber \\ 
&&\times 
\left[ 
F_{L' L'}(r_*, r_*) {\cal I}_{\ell_3 \ell_4}^{\ell_1 \ell_2}(L_1, L_3, L'; n, L)
+ F_{L' L}(r_*, r_*)  
{\cal J}_{\ell_3 \ell_4}^{\ell_1 \ell_2}(L_1, L_3, L'; n, L)
\right. \nonumber \\  
&& \left. \quad  
+ F_{L L'}(r_*, r_*)  
{\cal J}^{\ell_3 \ell_4}_{\ell_1 \ell_2}(L_3, L_1, L'; n, L) 
\right] ~, \label{eq:CMB_red_tris_app}  
\end{eqnarray}
where 
\begin{eqnarray}
R(\ell_1, L_1, \ell_2) \equiv \int_0^\infty r^2 dr 
\beta_{\ell_1 L_1}(r) 
\alpha_{\ell_2}(r) ~.  
\end{eqnarray}
This approximate formula enables us to calculate the trispectrum in the
whole $\ell$ space within a reasonable computational time. This approximation is justified, as the signal-to-noise of the trispectrum is dominated by soft limits in which $L$ is small \cite{Kogo:2006kh}. 
Using a scale-invariant curvature power spectrum, $\frac{k^3
P_\zeta(k)}{2\pi^2} = A_S$, we obtain $F_{LL'}(r_*, r_*)$ as
\begin{eqnarray}
F_{LL'}(r_*, r_*) = \frac{\pi^2}{2} A_S
\frac{  \Gamma(\frac{L+L'}{2})}
{\Gamma(\frac{L-L'+3}{2}) \Gamma(\frac{L'-L+3}{2})
\Gamma(\frac{L+L'+4}{2}) } ~.
\end{eqnarray}

Let us also derive eq.~(\ref{eq:CMB_red_tris_app}) from
eq.~(\ref{eq:CMB_red_tris}) using the Sachs-Wolfe approximation. In the
Sachs-Wolfe limit, the transfer function is given by ${\cal T}_{\ell}(k)
\to -\frac{1}{5} j_{\ell}(kr_*)$, and hence $\alpha_\ell(r) \to
-\frac{1}{5 r_*^2}\delta(r - r_*)$. Performing the $r$ and $r'$
integrals, we recover eq.~(\ref{eq:CMB_red_tris_app}) with $R(\ell_1,
L_1, \ell_2) \to \frac{1}{25} F_{\ell_1 L_1}(r_*, r_*)$. 

\section{Expected error bars on  $d_0$ and $d_2$}\label{sec:Fisher}

In this section, we calculate the expected 68\%~CL error bars on $d_n$
using the full-sky formalism. Let us define a Fisher matrix element for
$d_n$ as \cite{Hu:2001fa} 
\begin{eqnarray}
F_{n n'} \equiv \sum_{\ell_1 > \ell_2 > \ell_3 > \ell_4} \sum_L 
\frac{T_{\ell_3 \ell_4}^{\ell_1 \ell_2}(L, n) T_{\ell_3 \ell_4}^{\ell_1 \ell_2}(L, n')}{(2L + 1)C_{\ell_1} C_{\ell_2} C_{\ell_3} C_{\ell_4}} ~,
\end{eqnarray}
where $C_\ell$ is the temperature power spectrum. We shall consider an
ideal, noise-free, cosmic-variance limited experiment measuring
temperature anisotropy up to a maximum multipole of $\ell_{\rm max}$;
thus, $C_\ell$ contains the CMB only.

The trispectrum averaged over possible orientations of quadrilaterals,
$T_{\ell_3 \ell_4}^{\ell_1\ell_2}(L, n)$, is given by \cite{Hu:2001fa} 
\begin{eqnarray}
T_{\ell_3 \ell_4}^{\ell_1 \ell_2}(L,n) 
&=& P_{\ell_3 \ell_4}^{\ell_1 \ell_2}(L,n) + 
(2L+1) \sum_{L'} 
\left[ (-1)^{\ell_2 + \ell_3} 
 \left\{
  \begin{array}{ccc}
  \ell_1 & \ell_2 & L \\
  \ell_4 & \ell_3 & L' 
  \end{array}
 \right\}
P_{\ell_2 \ell_4}^{\ell_1 \ell_3}(L',n)  \right. \nonumber \\ 
&&\left.\qquad\qquad\qquad\qquad\qquad + (-1)^{L+L'} 
\left\{
  \begin{array}{ccc}
  \ell_1 & \ell_2 & L \\
  \ell_3 & \ell_4 & L' 
  \end{array}
 \right\}
P_{\ell_3 \ell_2}^{\ell_1 \ell_4}(L',n) \right], ~\label{eq:cmb_red_tris_form}
					   \end{eqnarray} 
with 
\begin{eqnarray}
P_{\ell_3 \ell_4}^{\ell_1 \ell_2}(L,n) &=& 2 t_{\ell_3 \ell_4}^{\ell_1 \ell_2}(L,n) 
+ 2 (-1)^{\ell_1 + \ell_2 + L} t_{\ell_3 \ell_4}^{\ell_2 \ell_1}(L,n) \nonumber \\ 
&&+ 2 (-1)^{\ell_3 + \ell_4 + L} t_{\ell_4 \ell_3}^{\ell_1 \ell_2}(L,n) 
+ 2 (-1)^{\ell_1 + \ell_2 + \ell_3 + \ell_4} t_{\ell_4 \ell_3}^{\ell_2 \ell_1}(L,n)~.
\end{eqnarray}

\begin{figure}
  \begin{center}
    \includegraphics[width =0.75\textwidth]{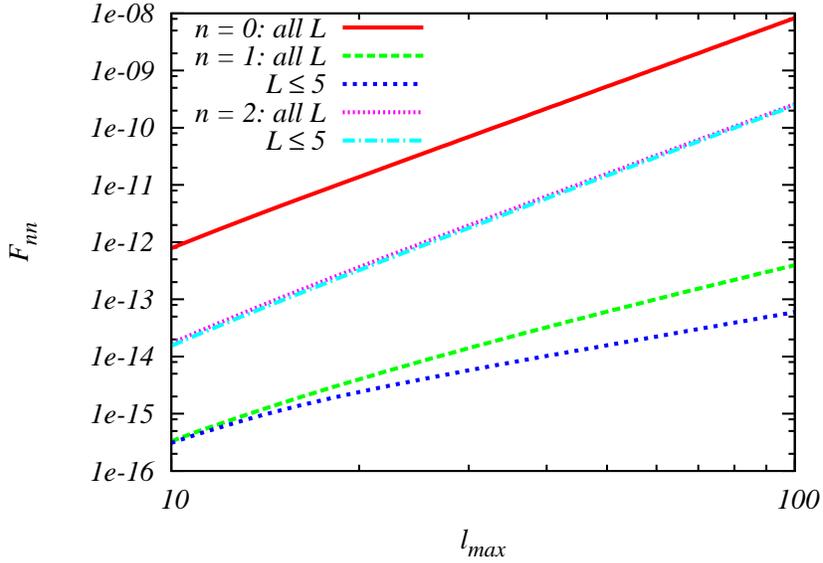}
  \end{center}
  \caption{Diagonal elements of the Fisher matrix, $F_{00}$, $F_{11}$,
 and $F_{22}$, computed using the Sachs-Wolfe approximation. The lines
 with ``all L'' use all $L$ in the summation of the Fisher matrix, while
 the lines with ``$L\le 5$'' for $n=1$ and 2 use only  $L\le 5$.}
  \label{fig:Fisher_SW}
\end{figure}

Figure~\ref{fig:Fisher_SW} shows the diagonal elements of the Fisher
matrix, $F_{00}$, $F_{11}$, and $F_{22}$, computed using the Sachs-Wolfe
approximation. We show the results from summation over all possible
diagonals, $L$, as well as those from summation over only soft limits,
$L\le 5$. We find that $F_{00}$ grows
as $\ell_{\rm max}^4$ in agreement with the previous work
\cite{Kogo:2006kh}, and $F_{22}$ also grows as  $\ell_{\rm max}^4$;
however, $F_{22}$ is smaller than $F_{00}$ by two orders
magnitude. Most of information of the trispectrum with $n=2$ is
contained in the soft limit, $L\le 5$, just like that with $n=0$. 
On the other hand, $F_{11}$ grows more slowly as $\ell_{\rm
max}^3$, implying that the error bar on $d_1$ would be too large to be
useful. We thus do not consider $d_1$ any further in this paper. 
Information of the trispectrum with $n=1$ is not completely
contained in the soft limit, and sizable contributions come from $L>5$.

\begin{figure}
  \begin{center}
    \includegraphics[width =0.75\textwidth]{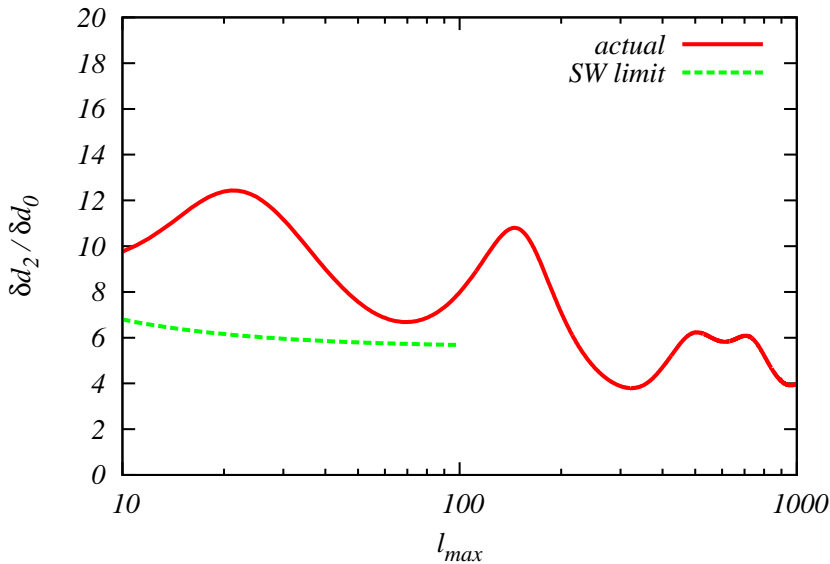}
  \end{center}
  \caption{Ratio of the expected error bars, $\delta d_2 / \delta d_0$. The solid line shows the results
 from eq.~(\ref{eq:CMB_red_tris_app})  with the full radiation transfer
 function, while the dashed line shows the Sachs-Wolfe approximation.}
  \label{fig:error_d2}
\end{figure}

We now calculate the expected error bars on $d_0$ and $d_2$ when they
are estimated jointly. We use
\begin{eqnarray}
{}^{(2)}F_{ij} =  \left(
  \begin{array}{ccc}
  F_{00} & F_{02} \\
  F_{02} & F_{22} 
  \end{array}
 \right)~,
\end{eqnarray}
to obtain
\begin{eqnarray}
(\delta d_0, \delta d_2) = 
\left(\sqrt{{}^{(2)}F_{11}^{-1}}, \sqrt{{}^{(2)}F_{22}^{-1}}\right).
\end{eqnarray}

In figure~\ref{fig:error_d2}, we show the ratio of $\delta d_2$ to
$\delta d_0$ as a function of $\ell_{\rm max}$. The error bar on $\delta
d_2$ improves slightly faster than 
that on $\delta d_0$ as $\ell_{\rm max}$ increases.
We find $\delta d_2/\delta d_0=4$ for $\ell_{\rm max}=1000$. We also find
that these two parameters are not correlated very much: the
cross-correlation coefficient, $F_{02} / \sqrt{F_{00} F_{22}}$, is as
small as 0.2. For $\ell_{\rm max} = 1000$, we find $(\delta d_0, \delta
d_2) = (105, 418)$. 
If a scaling relation $F_{00} \propto F_{22} \propto F_{02} \propto
\ell_{\rm max}^4$ holds for $\ell_{\rm max} > 1000$, the expected error
bars on $d_0$ and $d_2$ would become 
$(\delta d_0, \delta d_2) = (26, 105)$ for $\ell_{\rm max} = 2000$.
Recalling $d_0=\tau_{\rm NL}/6$, the error bar on $d_0$ we obtain here agrees
with that given in ref.~\cite{Kogo:2006kh}.

\section{Expected error bar on  $g_*$ from the CMB trispectrum}\label{sec:g}

The parameters of the power spectrum ($g_*$), the bispectrum ($c_n$),
and the trispectrum ($d_n$) can be related to each other once a model of
inflation is specified. Such a relation is a powerful probe of the
physics of inflation. In this section, we use inflation models with
a particular coupling between a scalar field driving inflation and a
vector field given by $I^2(\phi)F^2$ to relate the trispectrum
parameters with $g_*$. The trispectrum averaged over all possible
orientations of quadrilaterals is given by \cite{Shiraishi:2013vja}
\begin{eqnarray}
T_\zeta^{I^2 F^2} 
&\approx& 24 N^2 |g_*| 
\left[ (\hat{\bf k}_1 \cdot \hat{\bf k}_3 )^2 + (\hat{\bf k}_1 \cdot \hat{\bf k}_{12} )^2 + (\hat{\bf k}_{3} \cdot \hat{\bf k}_{12} )^2 - (\hat{\bf k}_1 \cdot \hat{\bf k}_3 )(\hat{\bf k}_1 \cdot \hat{\bf k}_{12} )(\hat{\bf k}_3 \cdot \hat{\bf k}_{12} ) \right] \nonumber \\ 
&&\times P_\zeta(k_1) P_\zeta(k_3) P_\zeta(k_{12})  + (23 {~\rm perm})~, \label{eq:zeta_tris_I2F2}
\end{eqnarray}
where $N\approx 60$ is the number of $e$-folds counted from the end of
inflation. The shape of this trispectrum is 99\% correlated with the trispectrum
without $(\hat{\bf k}_1 \cdot \hat{\bf k}_3 )(\hat{\bf k}_1 \cdot
\hat{\bf k}_{12} )(\hat{\bf k}_3 \cdot \hat{\bf k}_{12})$. Adjusting the
amplitude, we find that the following trispectrum is an excellent
approximation to eq.~\eqref{eq:zeta_tris_I2F2}:
\begin{eqnarray}
T_\zeta^{I^2 F^2} 
&\approx& 0.89 \times 24 N^2 |g_*| 
\left[ (\hat{\bf k}_1 \cdot \hat{\bf k}_3 )^2 + (\hat{\bf k}_1 \cdot \hat{\bf k}_{12} )^2 + (\hat{\bf k}_{3} \cdot \hat{\bf k}_{12} )^2 \right] 
\nonumber \\ 
&&\times 
 P_\zeta(k_1) P_\zeta(k_3) P_\zeta(k_{12}) + (23 {~\rm perm})~.
\label{eq:newshape}
\end{eqnarray}
The 99\% correlation means that
eqs.~\eqref{eq:zeta_tris_I2F2} and \eqref{eq:newshape} have nearly
identical shapes. The pre-factor 0.89 in eq.~\eqref{eq:newshape} is
 the ratio of the overall averages of trispectra computed
 numerically. One can understand 
 this ratio by angular-averaging the trispectra in soft limits, using
\cite{Fujita:2013qxa}: 
$(\hat{\bf k}_1 \cdot \hat{\bf k}_3)^2|_{\rm av} = 1/3$, $(\hat{\bf k}_1
\cdot \hat{\bf k}_{12})^2|_{\rm av} = (\hat{\bf k}_3 \cdot \hat{\bf
k}_{12})^2|_{\rm av} = 1/2$, and $(\hat{\bf k}_1 \cdot \hat{\bf k}_3)
(\hat{\bf k}_1 \cdot \hat{\bf k}_{12}) (\hat{\bf k}_3 \cdot \hat{\bf
k}_{12})|_{\rm av} = 1/6$. This gives a very similar value of 0.875.

Comparing the above expression with eq.~(\ref{eq:zeta_tris_def}) yields the
relationship between $d_0$, $d_2$ and $g_*$ as  
\begin{eqnarray}
d_0 = \frac{1}{2} d_2 \approx 0.89 \times 2880 \frac{|g_*|}{0.1} \left(\frac{N}{60}\right)^2 ~. \label{eq:rel_g}
\end{eqnarray} 

\begin{figure}
  \begin{center}
    \includegraphics[width =0.75\textwidth]{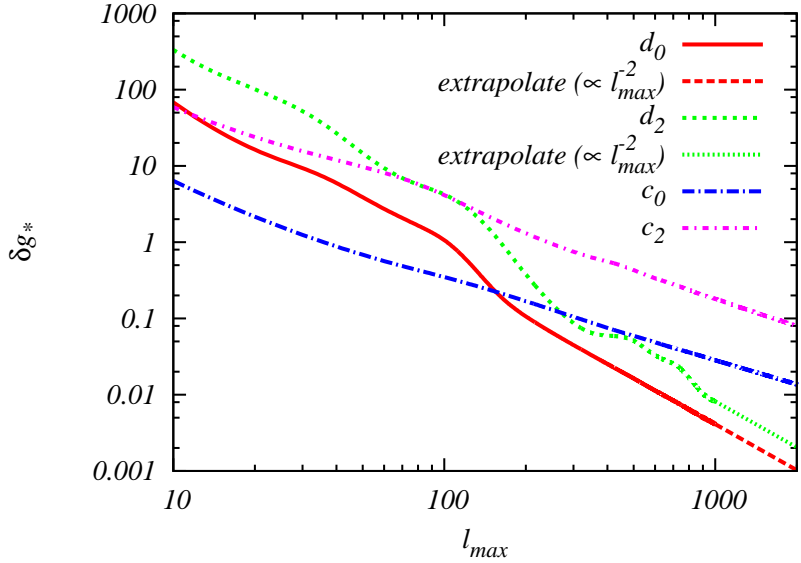}
  \end{center}
  \caption{Expected 68\%~CL error bars on $g_*$ from the bispectrum parameters
 ($c_0$ and $c_2$ in eq.~(\ref{eq:zeta_tris_def})) and the trispectrum
 parameters ($d_0$ and $d_2$), for $N=60$. The lines for the trispectrum in
 $\ell_{\rm max} > 1000$ are extrapolation.}
  \label{fig:error_g}
\end{figure}

In figure~\ref{fig:error_g}, we show the expected error bars on $g_*$
computed from those on $d_0$ and $d_2$ using eq.~\eqref{eq:rel_g}. We
show the results of the direct calculation of $\delta d_0$ and $\delta
d_2$ up to $\ell_{\rm max}=1000$, and use the extrapolation for
$1000<\ell_{\rm max}\le 2000$. For comparison, we also show the error
bars on $g_*$ from the bispectrum parameters using \cite{Shiraishi:2013vja}
\begin{eqnarray}
c_0 = 2 c_2 = 32 \frac{|g_*|}{0.1} \frac{N}{60} ~.
\end{eqnarray}

\begin{figure}
\centerline{
\includegraphics[width=\textwidth,angle=0]{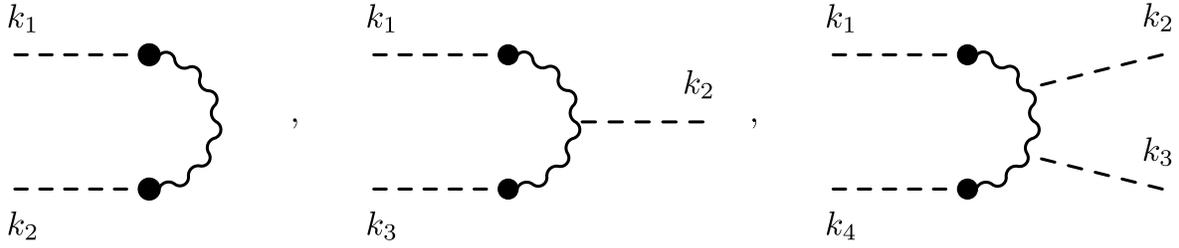}
}
\caption{Diagrams for the 2- (left), 3- (middle), and 4-point (right) functions of $\zeta \propto \delta \phi$ in the $I^2 \left( \phi \right) F^2$ model. The external dashed lines are $\delta \phi$ lines, while the internal propagators are $\delta {\bf E}$ lines. The labels denote the momentum of the external lines, which is taken to flow inside the diagram. In the power spectrum, ${\bf k}_1 + {\bf k}_2 = 0$, while in the other two diagrams we are interested in the soft-limit configurations ${\bf k}_1 + {\bf k}_2 \rightarrow 0$. A bullet denotes a mass insertion, namely a quadratic $\delta {\bf E} \; \delta \phi$ coupling proportional to the vector vacuum expectation value ${\bf E}_{\rm cl}$. 
}
\label{fig:diag}
\end{figure}

The trispectrum parameters are proportional to $|g_*|N^2$, whereas the
bispectrum parameters are proportional to $|g_*|N$. More generally, we have 
\begin{equation}
\langle \zeta^n \rangle \propto  \vert g_* \vert N^{n-2} \;,  
\label{scaling-zn}
\end{equation}
where $\zeta \propto - \frac{H}{\dot{\phi} }  \, \delta \phi $  in uniform density gauge. 
To understand this scaling, consider the diagrams shown in figure~\ref{fig:diag}, which represent the dominant contributions to $\langle \zeta^{2,3,4} \rangle$ arising from this interaction. By Taylor-expanding the $I^2 \left( \phi \right) F^2$ coupling in the inflaton perturbations $ \delta \phi \propto \zeta$, and by retaining only the linear terms,~\footnote{The higher order terms can be shown to give subdominant contributions \cite{Bartolo:2012sd}.
More in general, see ref.~\cite{Bartolo:2012sd} for the detailed computation of the power spectrum and bispectrum. 
The computation of the trispectrum is performed analogously \cite{Shiraishi:2013vja}.} we have the two interactions $H_1 \propto \int  d^3 x a^4 {\bf E}_{\rm cl} \cdot \delta {\bf E} \, \zeta$ and  $H_2 \propto \int d^3 x a^4 \delta {\bf E} \cdot \delta {\bf E} \, \zeta$. In this expression $H_i$ denotes a contribution to the interaction Hamiltonian, and $a$ is the scale factor ($a^4 = \sqrt{-g}$ in conformal time $\tau$). For each value of $n$, the diagram shown in the figure corresponds to the following terms in the in-in formalism computation 
\begin{equation}
\langle \zeta^n \left( \tau \right) \rangle 
\propto \left[ \prod_{i=1}^{n-1} \int d \tau_i \right] \left\langle \left[ \left[ \dots \left[ \zeta_0^n \left( \tau \right), H \left( \tau_1 \right) \right] , \dots \right]  , H \left( \tau_{n-1} \right) \right] \right\rangle \;, 
\label{zeta-n}
\end{equation} 
where $\zeta_0$ denotes the (``unperturbed'') curvature perturbation in the absence of the $I^2 \left( \phi \right) F^2$ term. We are interested in the correlators $\langle \zeta^n \rangle$ in the super-horizon regime. The integrals in eq.~(\ref{zeta-n}) are dominated by the regions in which also the fields arising from the vertices  are in the super-horizon regime \cite{Bartolo:2012sd}. Each interaction contains one $\zeta_0 \left( \tau_i \right) \propto \delta \phi \left( \tau_i \right)$ field which, once commuted with one of the external fields, gives $\left[ \zeta_0 \left( \tau \right) ,  \zeta_0 \left( \tau_i \right) \right] \propto \tau^3 - \tau_i^3$  \cite{Bartolo:2012sd}. These commutators, and the measure  $a^4 \left( \tau_i \right) \propto \frac{1}{\tau_i^4}$ in each vertex, are the only time-dependent contributions to the integrand in eq.~(\ref{zeta-n}), leading to     \cite{Bartolo:2012sd} 
\begin{equation}
\langle \zeta^n \left( \tau \right) \rangle \propto \prod_i^{n-1} \int^\tau \frac{d \tau_i }{\tau_i} \left( \tau^3 - \tau_i^3 \right) \propto N^{n-1}~.
\end{equation}

We thus see that the contribution to $\langle \zeta^n \rangle$ from the corresponding diagram in figure \ref{fig:diag} is $\propto E_{\rm cl}^2 \, N^{n-1}$. The diagram shown for the power spectrum ($n=2$ in this expression) adds up with the vacuum one, and provides the subdominant  quadrupole modulation $\propto \vert g_* \vert \propto  E_{\rm cl}^2 \, N$. Therefore, $\langle \zeta^n \rangle \propto E_{\rm cl}^2 \, N^{n-1} \propto \vert g_* \vert N^{n-2}$, as indicated in eq.~(\ref{scaling-zn}). It is also worth noting that each internal line in the diagram produces  in the final expression for $\langle \zeta^n \rangle $     a power spectrum which is function of the momentum carried on that line. For each given $n$, the diagram shown in the figure needs to be summed over with the diagrams obtained by permuting the position of the external lines. The diagrams shown in the figure  factor out a $P_\zeta \left( k_{12} \right)$, and are enhanced in the soft limit $k_{12} \rightarrow 0$. 

As a result of the scaling (\ref{scaling-zn}), if the error bars on
$c_n$ and $d_n$ are equal, the trispectrum is more sensitive to $g_*$
than the bispectrum by a factor of $N\approx 60$. In addition, we find
that the error bars on $g_*$ from the trispectrum decrease as 
$\delta g_*\propto \ell_{\rm max}^{-2}$, while those from the bispectrum
decrease more slowly as $\delta g_*\propto \ell_{\rm max}^{-1}$.
In reality, the error bars on the trispectrum parameters are much larger than
those on the bispectrum parameters for smaller $\ell_{\rm max}$; thus, we find
that the trispectrum yields smaller error bars on $g_*$ than the
bispectrum for $\ell_{\rm max}\gtrsim 200$.

For $\ell_{\rm max} = 2000$, we find $\delta g_* = 1.0 \times 10^{-3}$
and $2.0 \times 10^{-3}$ from the $d_0$ and $d_2$ measurements,
respectively. These error bars are an order of magnitude better than
those expected from the bispectrum measurements, and are comparable to
that expected from the power spectrum measurement for the same $\ell_{\rm max}$
(in the absence of systematic errors such as ellipticity of beams)
\cite{Pullen:2007tu}.

\section{Conclusion}\label{sec:conclusion}

Inflation models with anisotropic sources can create the perturbations
with a preferred direction, and yield distinct angular dependence 
not only in the power spectrum and bispectrum, but also in the trispectrum of
the CMB. Motivated by inflation models with $I^2(\phi)F^2$ coupling, we
have studied the observational consequence of the parametrized form of
the trispectrum given by eq.~(\ref{eq:zeta_tris_def}). 
The expected 68\%~CL error bars on the trispectrum parameters are $\delta d_0 =
26$ and $\delta d_2 = 105$ for a cosmic-variance-limited experiment
measuring temperature anisotropy up to $\ell_{\rm max} = 2000$. The
error bar on $d_1$ is too large to be useful.

Using the prediction of inflation models with $I^2(\phi)F^2$ coupling,
we derive the relationship between the trispectrum parameters and the
power spectrum parameter, $g_*$. We then find that the trispectrum
measurements can give competitive limit on $g_*$ reaching $\delta
g_*=10^{-3}$ for $\ell_{\rm max}=2000$, which is an order of
magnitude better than the expected limit from the bispectrum for the
same $\ell_{\rm max}$. This is
owing to two effects: the trispectrum parameters are proportional to
$|g_*|N^2$ whereas the bispectrum parameters are proportional to
$|g_*|N$; and the error bar on $g_*$ from the trispectrum decreases as
$\ell_{\rm max}^{-2}$ whereas that from the bispectrum decreases as
$\ell_{\rm max}^{-1}$. 

The signatures of broken rotational invariance in the power spectrum
\cite{Kim:2013gka} and the bispectrum \cite{Ade:2013ydc} have been
constrained by the temperature data of the {\it Planck} satellite. They
have yielded the limit on $g_*$ of order $10^{-2}$. This limit can be
improved further by using the trispectrum. The current limit on
$d_0$ from the {\it Planck} data, $d_0=\tau_{\rm NL}/6<470$ (95\%~CL)
\cite{Ade:2013ydc}, implies $|g_*|<0.02$ (95\%~CL), which is indeed
competitive. The other parameter, $d_2$, has not been constrained by the
data yet. Measurements of $d_0$ and $d_2$ from the full data of {\it
Planck} should yield the best limit on $g_*$ within the context of
inflation with $I^2(\phi)F^2$ coupling.

\acknowledgments
MS thanks Frederico Arroja for useful discussion on the shape correlator of the curvature trispectrum. 
MS was supported in part by a Grant-in-Aid for JSPS Research under Grant No.~25-573 and the ASI/INAF Agreement I/072/09/0 for the Planck LFI Activity of Phase E2. 
MP was supported in part from the DOE grant DE-FG02-94ER-40823 at the University of Minnesota.

\bibliography{paper}
\end{document}